\documentclass[submission,copyright,creativecommons]{eptcs}
\providecommand{\event}{NCMA 2022} 
\usepackage{verbatim}
\usepackage{iftex}

\usepackage[utf8x]{inputenc}

\ifpdf
  \usepackage{underscore}         % Only needed if you use pdflatex.
  \usepackage[T1]{fontenc}        % Recommended with pdflatex
\else
  \usepackage{breakurl}           % Not needed if you use pdflatex only.
\fi

\usepackage{amsfonts}
\usepackage{amssymb}
\usepackage{amsmath}
\usepackage[dvipsnames]{xcolor}
\usepackage{graphicx}
\usepackage{tikz}
\usetikzlibrary{positioning,arrows.meta,calc,fit}
\usepackage{enumitem}

\usepackage{amsthm}

\title{
Language and Intelligence, Artificial vs. Natural 
or
What Can and What Cannot AI Do with NL?
}

\def\titlerunning{Language and Intelligence, Artificial vs. Natural
or
What Can and What Cannot AI Do with NL?}

\author{Gyula Klima
  \institute{
  Department of Philosophy\\
  Fordham University}
  \email{klima@fordham.edu}
}
\def\authorrunning{Gyula Klima}

\begin{document}

\maketitle

\begin{abstract}
In this talk, I argue that there are certain pragmatic features of natural language (that I will call 'productivity' and 'malleability', on top of syntactical generativity and semantical compositionality), which are not only hard, but even impossible to capture in an artificial language used by an AI system, and the reason for this is to be found in certain deep, metaphysical differences between artificial and natural intelligence, accounting for the differences in their respective processes of concept-formation.
\end{abstract}

%\paragraph{Keywords:} Boolean networks, controllability, formal framework

%%%%%%%%%%%%%%%%%%%%%%%%%%%%%%%%%%%%

\section{Introduction: some simple definitions, and two theses}
%\label{sec:intro}

%%%%%%%%%%%%%%%%%%%%%%%%%%%%%%%%%%%%

More specifically, I will argue for two theses, one negative, one affirmative. The negative one will present what AI cannot do with natural languages that we humans can. The affirmative one will present what AI can do with natural languages we humans can. To better articulate these claims, let me first introduce some definitions. Some of these are common, some of these are not, but for the purposes of this talk I will just Humpty Dumptyishly use the terms defined here in accordance with the following definitions.

A sign is something that besides presenting itself also presents something (or some things) as its semantical value(s) to a cognitive subject. We can quickly note here that a semantical value of a sign can coincide with the sign itself, but then it is not presenting itself only as just a thing, but also as that which it is a sign of.

A language is a system of signs, consisting of a set of simple signs (called a vocabulary) and a set of complex signs (called well-formed expressions) generated by some rules of construction (called the grammar of the language).

\begin{enumerate}
\item A language has the syntactical feature of generativity if its grammar allows the generation of a potential infinity of well-formed expressions even from a finite vocabulary.
\item A language has the semantical feature of compositionality if the semantical values of its well-formed expressions are a function of the semantical values of its simple components. 
\item A language has the pragmatic feature of productivity if its users can produce new, both syntactically and semantically simple signs in its vocabulary (where a sign is both semantically and syntactically simple when it is not generated by generativity and its semantic values are not determined by compositionality).
\item Finally, a language has the pragmatic feature of malleability if its users can systematically reassign the semantic values of both its existing vocabulary and its expressions for the sake of their practical purposes at will.
\end{enumerate}
Based on these definitions or "meaning postulates," my two theses are as follows:
\begin{enumerate}[label=\Roman*.]
\item AI cannot use, let alone produce, a language that has all four features listed above. 
\item AI can use, and even produce, a language that has the first two features listed above.
\end{enumerate}
Perhaps, thesis II is not too surprising. After all, the first two features are quite obviously common features of both natural and artificial languages, the latter of which are AI's "daily bread," so to speak. Well, but would a however advanced AI system "get" my point about formal languages being their "daily bread" in the way you certainly did (understanding the allusion to the Lord's Prayer, along with the literal meaning of the phrase, as well as its intended meaning as referring to something, indeed, anything, that provides your everyday sustenance in your life and work)?

To be sure, this is not my argument for thesis I, rather it is just a preliminary indication of its possible implications. My main argument for thesis I is in fact more arcane: it is a metaphysical argument from Thomas Aquinas for the immateriality of the human intellect from its ability to produce new, simple universal concepts by abstraction. Obviously, to connect that metaphysical argument meaningfully to my negative thesis, I still have some more preliminary work to do, namely, I need first to present Aquinas' semantic conception, relating language, thought, and reality in a particular, and I'd say, particularly intriguing fashion.

%%%%%%%%%%%%%%%%%%%%%%%%%%%%%%%%%%%%

\section{Some bits and pieces of Aquinas' semantics and cognitive psychology}
%\label{sec:preliminaries}

%%%%%%%%%%%%%%%%%%%%%%%%%%%%%%%%%%%%

What I am about to present here is not the entirety of Aquinas' semantic conception. I dealt with that elsewhere, but now I present only the part that is relevant to his argument I am considering here, namely, the semantics of simple, universal categorematic terms, such as 'dog' 'tree,' 'human,' etc., in all ten Aristotelian categories.

Perhaps, the easiest way to do this is by using a diagram I used in my Stanford Encyclopedia article on \underline{the medieval problem of universals}.\\

\hspace*{-0.7cm}
\includegraphics[width=1\textwidth]{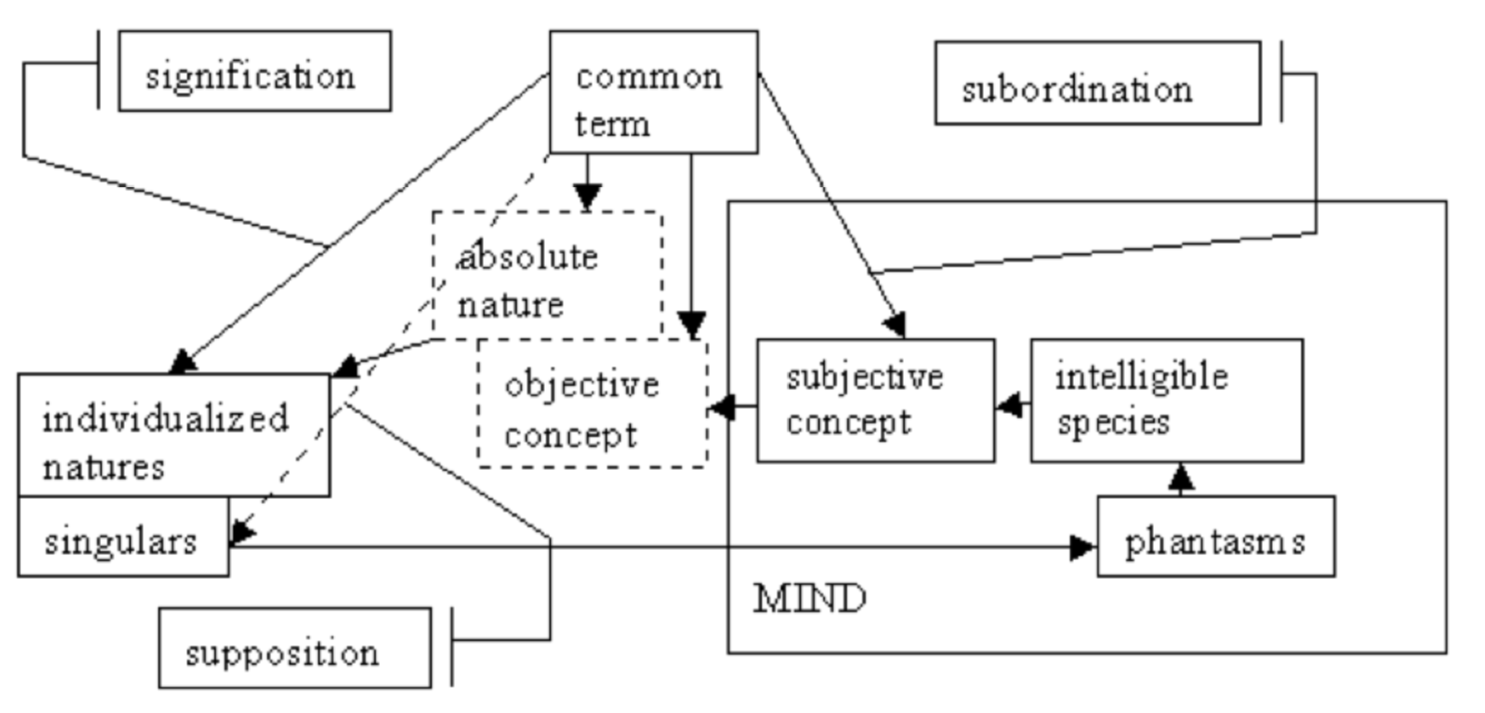}

In this diagram, the arrows pointing from the common term indicate semantic relations (in particular, subordination, which determines that all conventional semantic features of our linguistic terms are inherited from the semantic features of the concepts to which they are subordinated), whereas the arrows pointing from singulars to phantasms (the singular sensory representations of singular things), and from phantasms to concepts (the universal representations of singular things actually used in the mind) indicate the flow of information.

For instance, if I say, 'This is round, white, and cold', pointing to a snowball, then the predicates of this sentence will be taken to signify its round shape, its white color, and its cold temperature. But of course, these terms signify not only the individualized qualities of this particular snowball, because they signify these qualities universally, in abstraction from the individuating conditions of their subject, determining that they apply to this thing here and now. Thus, what we are going to be concerned with in particular is the short arrow pointing from phantasms to intelligible species, the first universal representations of the human mind deposited in intellectual memory for use in acts of concept formation when these concepts are needed in the formation of thoughts involving universal representations of singulars. That short arrow indicates the process of intellectual abstraction, the production of intelligible species. In the interest of time, let me refer you to the Stanford Encyclopedia article for the rest of the details of this diagram, and let us focus here only on what is relevant in it to Aquinas' argument.

For a better understanding of Aquinas' point, after presenting a brief sketch of the argument itself, I am going to discuss it by responding on Aquinas' behalf to what I take to be the most serious objection to it, raised by the Parisian nominalist philosopher of the $14^\textnormal{th}$ century, John Buridan. I would say, however, that Buridan's objection would be typically raised also by our contemporaries, as it seems to be so plausible in terms of our contemporary intuitions. This should not come as a surprise, though, given that those intuitions have actually been shaped in a long and complex historical process by Buridan's and other nominalists' logical insights. But I cannot and will not go into this historical issue here. So, let us see first Aquinas' argument, then Buridan's objection, and then my response, pointing out an important conceptual conflation concerning universal representation, which will turn out to be directly relevant to our original question of what an AI system cannot do that a human intelligence can, and why.

%%%%%%%%%%%%%%%%%%%%%%%%%%%%%%%%%%%%

\section{Aquinas' argument from the universality of concepts}

%%%%%%%%%%%%%%%%%%%%%%%%%%%%%%%%%%%%

The main claim of Aquinas' argument is that the universal concepts of our understanding cannot be received in a material medium because their universality is achieved precisely by their being abstracted from matter.

To be sure, the universal, abstract mode of representation of the concepts of the intellect alone cannot guarantee their immateriality -- after all, we are all familiar with material universal signs, such as the words we utter or write. However, these universal symbols can have their universal representative function only because they are subordinated to the primarily universal concepts of our understanding. So, their derivative, conventional universality need not entail any ontological constraints upon their nature. Such a constraint may, however, be entailed by the primarily
universal representative function of the concepts of understanding, which are formed as a result of the natural causality of sensible objects on the senses and the consequent further processing of the information thus obtained from the senses by the intellect.

The question then is whether, and if so, why the primarily universal mode of representation of the concepts of the understanding formed as a result of the causality of sensible objects should entail the immateriality of these concepts, in the sense that the subject in which they are received, the intellect, cannot be material.

The main idea is the following. The senses represent singulars in their singularity because they necessarily represent the sensible features of material objects together with the material individuating conditions of these features, namely, the spatio-temporal dimensions determining their principle of individuation, the designated matter of these objects, that is, the matter of these objects contained under their dimensions here and now. The reason why this is necessary is that the causally active sensible features of sensible objects necessarily exercise their causality on the senses under these determinate dimensions, and so these sensible features are necessarily encoded by the senses as determined by these dimensions. Therefore, what encodes these spatio-temporal features in the senses are some corresponding spatio-temporal features of the sense organs because it is precisely these features of the sense organs that are impacted by the corresponding features of their objects. For example, the spatial arrangement of distinct patches of color in my visual field is encoded by the spatial pattern of receptors firing in the retina of my eyes, and similar considerations apply to the other senses. In the process of abstraction, therefore, the intellect has to "cut out" precisely this part of the code preserved in the phantasms. So, it has to form the concepts encoding the universal information contained in a vast number of different phantasms in a medium that will not encode the information about the singularity of singulars represented by the phantasms. Yet, it will encode that part of the information that is invariably common, because essential, in this massive "database" of various singular experiences under variable circumstances such as time and location and other coincidental circumstances and features.

But then, since what encodes this circumstantial information in the phantasms is the spatio-temporal features of the organs in which they are received, the medium in which the universal representations are formed must be something that does not have such spatio-temporal features, i.e., something that does not have its own dimensions, which can only be a thing that is immaterial; therefore, the intellect receiving these representations has to be immaterial. 

As can be seen, the most important idea in this argument is that the singularity of representation is necessarily tied to the materiality of representations. Sensory representation is necessarily singular because it is material, since the singularity of the information in sensory representation is encoded precisely by the material features of sensory representations, namely, the spatio-temporal features of the sense organs (including relevant parts of the brain) that are modified according to the spatio-temporal features of sensory objects, which determine their singularity. But if this much is acceptable, then the argument can indeed establish its desired conclusion.

All in all, the argument boils down to just the following (instantiated) modus tollens:
\begin{enumerate}
\item If a natural (as opposed to conventional) representation is material, then it is singular
\item (Some of) the intellect's natural representation is not singular (since it is universal)
\item Therefore, (some of) the intellect's natural representation is not material
\end{enumerate}
Once we reduce Aquinas' rather complex original argument to this simple modus tollens, then it should become clear that its crucial part is the first premise, namely, the alleged implication moving from the materiality of a natural representation to its singularity, by natural (as opposed to logical) necessity. So, it is not surprising that it is precisely this implication that Buridan targets in his criticism.\footnote{For a more detailed presentation of and references to Aquinas' argument, see \cite{kl2018}}

%%%%%%%%%%%%%%%%%%%%%%%%%%%%%%%%%%%%

\section{Buridan on Aquinas' argument}

%%%%%%%%%%%%%%%%%%%%%%%%%%%%%%%%%%%%

Buridan begins his discussion with the claim that the position we find in Aquinas' Commentary on De Anima is wrong:
\small
\begin{itemize}
\item[]
... it has seemed to some people that sense lacks the nature for cognizing universally, although it does have it for cognizing singularly and determinately, because it has extension and a determinate location in a corporeal organ. It is the other way around with the intellect, however. Because it is immaterial and unextended, not determining for itself a location in a corporeal organ, it does possess the nature for understanding universally and not singularly. [...] But this opinion seems defective.\footnote{Buridan, QDA3 q. 8, nn. 18-19. in \cite{kletal22}}
\end{itemize}
\normalsize
There are two passages in which Buridan explicitly takes on the task of refuting (1). First, he writes:
\small
\begin{itemize}
\item[]
... an extended power, such as a horse's appetite, is indeed brought to its object in a universal way. For the thirsty horse desires water, and not determinately this water to the exclusion of that, but indifferently, any water at all. Thus, it drinks whichever it finds first.\footnote{QDA3 q. 3, n. 30. in \cite{kletal22}}
\end{itemize}
\normalsize
Likewise, in q. 8, the idea is the indifferent, and hence apparently universal natural desire of the agent, indeed, even of an inanimate agent, which does not require an immaterial subject for its universality:
\small
\begin{itemize}
\item[]
Furthermore, it is apparent that a material and extended power is properly brought to bear on its object in a universal way, for the appetite of a horse in the form of hunger or thirst is not singularly for this sack of oats or that water, but for any, indifferently, which is why it would take whichever it finds first. And the natural intention or appetite of fire for heating is not related to this or that heatable thing in a singular way, but indifferently, to anything it can heat. Thus, it would heat whatever is put to it.\footnote{QDA3 q. 8, n. 24. in \cite{kletal22}}
\end{itemize}
\normalsize
However, despite his criticism of Aquinas' implication, Buridan does agree with him on the claim that the universality of a cognitive act is due to the abstractive ability of the cognitive power in question, and the singularity of a cognitive act is due to it carrying distinctive information about singulars as such, from which it is unable to abstract:
\small
\begin{itemize}
\item[]
... although an exterior sense cognizes Socrates, or whiteness, or white (color), this is only via a species representing it confusedly with the substance, the whiteness, the size, and location, as he appears in the prospect of the person cognizing him. And the power of sense cannot sort out the confusion if it is unable abstract the species of the substance, the whiteness, the size, and location from each other, and so it can only perceive substance, or whiteness, or a white thing in the manner of something existing in its prospect. Thus, it can cognize the aforementioned objects only singularly.\footnote{QDA3 q. 8, n. 28. in \cite{kletal22}}
\end{itemize}
\normalsize
Indeed, since he thus ties the singularity of cognition to its distinctive representational content about individuals, Buridan argues even further that the singularity of representation is inherited by all those sensory powers that do not abstract from this distinctive, singularizing information:
\small
\begin{itemize}
\item[]
Again, since\footnote{My emendation ad sensum, for the text's 'even if' [etsi].} the common sense receives [the sensible] species from the external senses with this sort of confusion and cannot resolve the confusion, it necessarily apprehends in a singular manner. That is why we judge in dreams that this or that appears to us to exist, or to be here or there.\footnote{QDA3 q. 8, n. 29. in \cite{kletal22}}
\end{itemize}
\normalsize
Thus, just like Aquinas, Buridan clearly ties the singularity of sensory representation to its distinctive information content. Nevertheless, in his criticism of Aquinas' position, he also allows that some material, sensory powers, such as sensory appetite, can relate to their objects in an indifferent, non-singular manner, which presupposes some sort of indifferent, apparently universal representation of their objects. But is that really the sort of genuinely universal representation we have in intellectual cognition? Indeed, is Buridan's criticism consistent with his position on the reason for the singularity of sensory representation based on the different location of singulars in one's sensory field?

%%%%%%%%%%%%%%%%%%%%%%%%%%%%%%%%%%%%

\section{My response to Buridan on Aquinas' behalf}

%%%%%%%%%%%%%%%%%%%%%%%%%%%%%%%%%%%%

Answering first this second question, we should know that Aquinas provides a very compelling reason why location can have a crucial, naturally distinctive role in singular cognition, while discussing the sensory cognition of common sensibilia (i.e., sensible qualities that are cognized by several senses, such as shape, size, position, etc., as opposed to proper sensibilia that are cognized only by one sense):
\small
\begin{itemize}
\item[]
... there are objects which differentiate sensation with respect not to the kind [species] of the agent, but to the mode of its activity. For as sensible qualities affect the senses corporeally and locally, they do so in diverse ways if they are qualities of large or small bodies or are diversely situated, i.e., near, or far, or in the same place or in diverse places. And it is in this way that the common sensibles differentiate sensation.
\footnote{\textit{Sentencia De anima,} lib. 2 l. 13 n. 12. in \cite{kletal22}}
\end{itemize}
\normalsize
So, common sensibilia are the necessary spatio-temporal determinations of all proper sensibilia. The proper sensibilia, in turn, are the individualized sensible qualities of material individuals that the external senses are specifically attuned to be affected by, such as color, sound, smell, taste, texture and temperature. But these proper sensibilia are individualized by their spatio-temporal determinations here and now, the common sensibilia. The cognition of common sensibilia, therefore, provides precisely that distinctive sensory information that singularizes the cognition of individualized sensible qualities, presenting the singulars having these qualities qua the singulars actually affecting the senses here and now.

However, if the external senses receive this distinctive, singular information about the individuating spatio-temporal conditions of their objects on account of receiving the causal impact of these objects through their own spatio-temporal features, then this seems to establish the implication Buridan argued against in the first place -- that the materiality of a cognitive power entails the singularity of its cognitive act. For if sensory representation is singular precisely because it represents its object in a material fashion, encoding the distinctive, singular information about the object by its own material features, then this means that sensory representation is singular because it is material, i.e., its materiality implies its singularity.

Indeed, if he is successful in establishing this much, then, pace Buridan, Aquinas does have a good argument for the immateriality of the intellect. For the contrapositive of this implication, namely, that the non-singularity of its cognitive act entails the immateriality of a cognitive power, together with the fact that the intellect does have some non-singular, i.e., genuinely universal cognitive acts, establishes the immateriality of the intellect. Thus, when Buridan endorsed the claim that the singularity of sensory cognition is due to it carrying distinctive singular information that it cannot abstract from, he did not move too far away from Aquinas' position. Aquinas just made the further, quite plausible, claim that the encoding of this distinctive singular information in the senses is due to their material character, as they receive the localized, spatio-temporal, causal impact of material singulars in a similarly localized, spatio-temporal fashion. But then, it seems, Aquinas is quite entitled to his further conclusion concerning the immateriality of the intellect.

Furthermore, in response to my first question at the end of the previous section, we should note that Buridan's objection to Aquinas' proof rests on an unjustifiable conflation between the mere indifference of singular representation obtained by the senses and true universality of representation; thus, we should be really careful in separating the two.

For take the snapshot of an egg. That picture, despite the fact that it does not carry enough distinctive information to tell (from that information content alone) which particular egg it is the picture of, still carries information only about that particular egg; so, it is not really a universal representation of all eggs, it is a singular representation only of that particular egg. Indeed, it does not carry any information about a marble egg of the same shape and color, even if a snapshot of that fake egg might be indistinguishable from that of the genuine egg. To be sure, this still does not mean that the two pictures are absolutely indistinguishable: after all, based on their metadata (time and geolocation), we could tell with precision which is the picture of which object. So, these apparently indifferently representing images are not universal representations of many or even all eggs; they are just singular, yet in their direct information content (as opposed to their metadata) non-distinctive representations of their singular objects.

By contrast, a genuinely universal representation of a genuine egg is the result of abstracting precisely that form of eggs in general which comes from innumerable experiences of all sorts of singular eggs not only in a single human life, but over generations of human experiences, so firmly established that allows it to be encoded in all varieties of human languages, eventually even allowing a scientific investigation to establish its articulation in a genuine, scientific, quidditative definition, to the exclusion of marble eggs merely resembling in shape and color.

To be sure, one might object: don't I also see a common form, say the common form of oval shape, in the indifferent image of an egg?

I would respond that no, you don't. A properly universal representation of that shape would be the mathematical formula of an ellipsoid shape, articulating our pre-theoretical, yet genuinely universal concept of the same, which carries not only non-distinctive information about this particular egg that the snapshot does, but also distinctive, precise, universal information about all possible egg-shapes.\footnote{https://www.math.net/ellipsoid}
This is why I have kept emphasizing in my previous writings touching on this subject that the process of abstraction, pace Locke,\footnote{For the contrast with Locke and British Empiricism in general, see \cite{kl04}} results not in a mere loss of information, but in the active sorting out of accidental, coincidental, irrelevant information, from a massive database of singulars to the effect of gaining genuinely universal, essential information, representing all possible singulars, whether in the category of substances or accidents.\footnote{Cf. \cite{kl11}}
But couldn't this sort of abstraction still be realized in the same material medium (say, the brain) as the processing of incoming sensory information?

To be sure, operating on their material carriers, the originally singular representations can be rendered more and more indifferent, by stripping them of more and more explicit or implicit singularizing information, giving them the semblance of universal representations. Yet, by this simple process of mere loss of information they are rendered just more and more non-distinctive, yet not genuinely universal. Genuine, naturally produced universal representations, namely, the intelligible species, are produced by abstraction, which is not only a process of losing distinctive, singularizing information, but at the same time it is the process of gaining universal information carried by all singular representations of the same kind, so that these universal representations also cover singulars of the same kind we have not experienced, of which we have no singular representations at all.

What is needed for this is an agent outside the causal chains of merely manipulating the material carriers of singular information (rendered singular by their material causal histories), which is outside of these causal chains in the sense that it is capable of grasping that common, intelligible content that applies to a potential infinity of entities of the same kind, never represented as such by any of the representations of singulars but carried by them all. This is why the resulting truly universal representation cannot be material: it cannot have this information content resulting from the further material manipulation of the material carriers of singular information, which are singular precisely because of their material causal histories.

So, making the distinction between a merely non-distinctive, singular representation and a genuinely universal representation effectively defuses Buridan's objection: just because a horse may quench its thirst in search of any bucket of water, it doesn't mean it has a universal concept of water any more than the fact that any bucket of water may put out a small fire requires that water has a universal concept of fire. The indifference of action in singular causal relations primarily dependent on the kinds of agent and patient is part and parcel of the regularities of nature, which of course does not require or presuppose any sort of awareness in any of them. Therefore, when Buridan is talking about the horse's indifferent desire toward any bucket of water to quench its thirst, it does not require on the horse's part any sort of genuinely universal representation: the horse is just indifferently reacting to the singular representations of this or that bucket, equally good for satisfying its thirst. But Aquinas' argument is about genuinely non-singular, universal representations of the intellect; so, Buridan's objection, conflating genuinely universal representations with merely non-distinctive singular ones, simply misses its target.

%%%%%%%%%%%%%%%%%%%%%%%%%%%%%%%%%%%%

\section{Conclusion: the implications of Aquinas' argument for NL and AI}

%%%%%%%%%%%%%%%%%%%%%%%%%%%%%%%%%%%%

So, let me conclude with a brief summary of this discussion, and some of its further implications concerning our original questions about what AI can and what it cannot do with NL. 

The nominalist conflation of true universality with mere indifference of a singular representation does in the end totally undermine Buridan's criticism: just because the snapshot of an egg does not give you distinctive information about this egg distinguishing it from any other egg of the same size and shape, it doesn't mean that the snapshot is a universal representation of eggs in general. In fact, the picture still carries information only about the same egg you took the picture of, except, based on the information content of the picture alone (to the exclusion of its metadata), you cannot determine which one it pictures (although, based on the broader context of taking the picture encoded in its metadata, it can still quite easily be determined). 

Thus, Buridan seems to be implicitly committed to the implication he explicitly rejects, which renders his attempted refutation ineffective. But then, if this is correct, we may safely conclude that as far as Buridan's criticism is concerned, Aquinas may just be right.

However, if Aquinas is indeed right, then his immateriality thesis has far-reaching consequences even for contemporary AI research. Given that any computer we shall ever make will process information in its material medium, the inevitable conclusion is this: if Aquinas' main thesis is right, then all the information any AI machine processes can only be secondarily universal, riding on the universality of our primarily universal, human concepts, which can only be produced in the immaterial medium of the human mind. Thus, an AI system will never be creative in the way human intelligence is. An AI system will never form new concepts, it can only re-process (perhaps, much faster, and never tiring, but those are other issues) the concepts produced by the genuinely creative, because immaterial human intelligence.

To be sure, this metaphysical limit on what AI can do with language may seem to be rather limited itself. After all, Aquinas' argument concerns only the natural, pre-linguistic process of the formation of both syntactically and semantically simple categorematic concepts that would populate Aristotle's system of categories. But the formation of these concepts is not what we would usually associate with human creativity but would rather regard it as characteristic of a necessary stage of children's developmental psychology, and rightly so. However, if Aquinas is indeed right, then his position can give us a clue about why a healthy human baby and a healthy chimpanzee baby raised in the same environment will develop so differently, especially with regard to their abilities of language learning and use. For on Aquinas' position, a human baby, having an immaterial intellect, can genuinely "produce" such immaterial concepts a chimpanzee baby is metaphysically incapable of producing. And this would be the metaphysical basis of the feature of natural languages I dubbed 'productivity.'

But, even further, Aquinas' position may also give us a clue about processes of secondary, post-linguistic abstractions, in terms of analogical and metaphorical, scientific, and poetic concept formation in interaction not only with one's sensible environment, but also with one's language and linguistic community.

For every word of our languages carries with it a palimpsest of concepts formed by the abstractions of generations of its users, recorded in its etymology. And this is the idea behind the fourth, pragmatic feature of natural languages I named 'malleability.' For our universal categorematic concepts we form in a natural process of abstraction are just those primordial simple concepts we can gain from our personal experience, which would be pretty much like the conceptual vocabulary of a feral child. But a child raised in a normal social environment is immediately exposed to the concepts encoded in the language of their environment. Thus, children's natural intelligence is going to form not only the primary primitive concepts based on their own personal experiences, but also the concepts of others, abstracted by generations and handed down to them through language. But this process of concept acquisition through language is not a mere copying process. 

Every user of a language has the ability to modify the use of conventionally significant words by subordinating them to their concepts acquired and modified by forming analogies and metaphors in a process of secondary abstraction. Take for instance the word 'malleability' as I am using it in this lecture. Perhaps, you are not even aware of the fact that at the etymological root of this English word is the Latin word 'malleus,' meaning hammer. So, strictly speaking, what is malleable is something that can be shaped and reshaped by a hammer. Yet, the way I am using it, the word is not used in its strict sense: of course, the concepts our words are subordinated to are not the kinds of things that can be reshaped by a hammer. But you had no difficulty in understanding my idea that our concepts,  and thus the usage of the words they are subordinated to are changeable, based on the analogy that just as malleable materials can be reshaped by a hammer, so our concepts can be modified in their representative functions by our minds, which accounts for the fact that our existing words getting subordinated to these modified concepts have this feature of 'malleability', of changing usage, endowing them with new meanings. But it is precisely this conceptual malleability that underlies what we ordinarily call "creative" uses of language whether in science, philosophy, or theology, using inventive analogies to introduce new concepts by modifying old ones, or in poetry and rhetoric, using metaphors for more powerful expression of our ideas that move both our intellect and imagination.

But if this sort of creative use of language is based on the secondary processes of abstraction that presuppose the primary process which by Aquinas' lights cannot take place in a material medium, but only in the immaterial medium of the human intellect, then it is unavailable to the material medium of an AI system. Nevertheless, the creative human mind will still be able to endow such systems with its already produced and further modified concepts, as long as it is capable of reproducing the semantic features of the corresponding human concepts in the semantic features of the words the AI system uses, exploiting those features of human languages that is available to such systems, namely, generativity and compositionality.%\\\\

%\noindent
%Gyula Klima\\
%Department of Philosophy\\
%Fordham University\\
%\underline{klima@fordham.edu}

%%%%%%%%%%%%%%%%%%%%%%%%%%%%%%%%%%%%

\bibliographystyle{eptcs}
\bibliography{refs}

\begin{thebibliography}{1}
\providecommand{\bibitemdeclare}[2]{}
\providecommand{\surnamestart}{}
\providecommand{\surnameend}{}
\providecommand{\urlprefix}{Available at }
\providecommand{\url}[1]{\texttt{#1}}
\providecommand{\href}[2]{\texttt{#2}}
\providecommand{\urlalt}[2]{\href{#1}{#2}}
\providecommand{\doi}[1]{doi:\urlalt{https://doi.org/#1}{#1}}
\providecommand{\eprint}[1]{arXiv:\urlalt{https://arxiv.org/abs/#1}{#1}}
\providecommand{\bibinfo}[2]{#2}

\bibitemdeclare{article}{kl2018}
\bibitem{kl2018}
\bibinfo{author}{Gyula \surnamestart Klima\surnameend} (\bibinfo{year}{2018}):
  \emph{\bibinfo{title}{Aquinas’ Balancing Act: Balancing the Soul Between
  the Realms of Matter and Pure Spirit}}.
\newblock {\slshape \bibinfo{journal}{Bochumer Philosophisches Jahrbuch für
  Antike und Mittelalter}} \bibinfo{volume}{21}(\bibinfo{number}{1}), pp.
  \bibinfo{pages}{29--48}, \doi{10.1075/bpjam.00022.kli}.
\newblock
  \urlprefix\url{https://www.jbe-platform.com/content/journals/10.1075/bpjam.00022.kli}.

\bibitemdeclare{article}{kl11}
\bibitem{kl11}
\bibinfo{author}{Gyula \surnamestart Klima\surnameend} (\bibinfo{year}{2018}):
  \emph{\bibinfo{title}{Aquinas’ Balancing Act: Balancing the Soul Between
  the Realms of Matter and Pure Spirit}}.
\newblock {\slshape \bibinfo{journal}{Bochumer Philosophisches Jahrbuch für
  Antike und Mittelalter}} \bibinfo{volume}{21}(\bibinfo{number}{1}), pp.
  \bibinfo{pages}{29--48}, \doi{10.1075/bpjam.00022.kli}.
\newblock
  \urlprefix\url{https://www.jbe-platform.com/content/journals/10.1075/bpjam.00022.kli}.

\bibitemdeclare{incollection}{kl04}
\bibitem{kl04}
\bibinfo{author}{Gyula \surnamestart Klima\surnameend},
  \bibinfo{author}{Peter~G. \surnamestart Sobol\surnameend},
  \bibinfo{author}{Peter \surnamestart Hartman\surnameend} \&
  \bibinfo{author}{Jack \surnamestart Zupko\surnameend} (\bibinfo{year}{2004}):
  \emph{\bibinfo{title}{John Buridan’s Questions on Aristotle’s De Anima
  – Iohannis Buridani Quaestiones in Aristotelis De Anima}}.
\newblock In \bibinfo{editor}{R.~L. \surnamestart Friedmann\surnameend} \&
  \bibinfo{editor}{S~\surnamestart Ebbesen\surnameend}, editors: {\slshape
  \bibinfo{booktitle}{John Buridan and Beyond: Topics in the Language Sciences
  1300-1700}}, \bibinfo{publisher}{Royal Danish academy of sciences and
  letters}, \bibinfo{address}{Copenhagen}, pp. \bibinfo{pages}{17--32}.

\bibitemdeclare{book}{kletal22}
\bibitem{kletal22}
\bibinfo{author}{Gyula \surnamestart Klima\surnameend},
  \bibinfo{author}{Peter~G. \surnamestart Sobol\surnameend},
  \bibinfo{author}{Peter \surnamestart Hartman\surnameend} \&
  \bibinfo{author}{Jack \surnamestart Zupko\surnameend} (\bibinfo{year}{2022}):
  \emph{\bibinfo{title}{John Buridan’s Questions on Aristotle’s De Anima
  – Iohannis Buridani Quaestiones in Aristotelis De Anima}}.
\newblock \bibinfo{publisher}{Springer Cham}, \bibinfo{address}{Switzerland}.

\end{thebibliography}

%%%%%%%%%%%%%%%%%%%%%%%%%%%%%%%%%%%%

\end{document}